\documentclass[%
 reprint,
superscriptaddress,
 amsmath,amssymb,
 aps,
]{revtex4-1}

\usepackage{graphicx}
\usepackage{dcolumn}
\usepackage{bm}
\usepackage{mathtools}
\usepackage{gensymb}
\usepackage{amsmath}
\usepackage{qcircuit}
\usepackage{dsfont,soul,xcolor} 

\DeclarePairedDelimiter\ket{\lvert}{\rangle}

\usepackage{xr}
\makeatletter
\newcommand*{\addFileDependency}[1]{
  \typeout{(#1)}
  \@addtofilelist{#1}
  \IfFileExists{#1}{}{\typeout{No file #1.}}
}
\makeatother

\newcommand*{\myexternaldocument}[1]{%
    \externaldocument{#1}%
    \addFileDependency{#1.tex}%
    \addFileDependency{#1.aux}%
}

\myexternaldocument{supplement}

\begin{document}


\title{Indefinite Causal Orders from Superpositions in Time}
\author{David Felce}
\affiliation{%
Clarendon Laboratory, Department of Physics, University of Oxford, Oxford OX1 3PU, United Kingdom
}%
\email{david.felce@physics.ox.ac.uk}
\author{Nicetu Tibau Vidal}
\affiliation{%
Clarendon Laboratory, Department of Physics, University of Oxford, Oxford OX1 3PU, United Kingdom
}%
\author{Vlatko Vedral}
\affiliation{%
Clarendon Laboratory, Department of Physics, University of Oxford, Oxford OX1 3PU, United Kingdom
}%
\affiliation{
Centre for Quantum Technologies, National University of Singapore, Block S15, 3 Science Drive 2, Singapore
}%
\affiliation{Department of Physics, National University of Singapore, Science Drive 3, Blk S12, Level 2, Singapore 1175512}
\author{Eduardo O. Dias}
\affiliation{%
Departamento de Fisica, Universidade Federal de Pernambuco, Recife, Pernambuco 50670-901, Brazil
}%




\begin{abstract}
Treating the time of an event as a quantum variable, we derive a scheme in which superpositions in time are used to perform operations in an indefinite causal order. We use some aspects of a recently developed space-time-symmetric formalism of events. We propose a specific implementation of the scheme and recover the Quantum SWITCH, where quantum operations are performed in an order which is entangled with the state of a control qubit. Our scheme does not rely on any exotic quantum gravitational effect, but instead on phenomena which are naturally fuzzy in time, such as the decay of an excited atom.

\end{abstract}

\maketitle


\paragraph*{Introduction---}
A growing literature has arisen describing events that take place without a well defined causal structure in quantum mechanics \citep{Hardy_2007,Ore,Giacomini_2016,PhysRevA.88.022318,Colnaghi2012,PhysRevLett.113.250402,PhysRevLett.120.120502,2018arXiv180906655S,chiribella2018indefinite,2019arXiv190201807P,2018arXiv181207508M,Zhao:19,PhysRevLett.125.070603,Procopio2015,PhysRevLett.121.090503,goswami2018communicating,PhysRevLett.122.120504,PhysRevLett.124.030502}. On the one hand, these works are motivated by the hypothesis that an eventual quantum theory of gravity might predict superpositions of spacetime geometries in which the causal relationship between events is itself in a superposition \citep{Hardy_2007}. On the other hand, it has been argued that indefinite causal orders (ICOs) can be found already within traditional quantum mechanics, without the need for superpostions of spacetime geometries \citep{PhysRevLett.113.250402,Procopio2015,PhysRevLett.121.090503,goswami2018communicating,PhysRevLett.122.120504,PhysRevLett.124.030502}.


An example of an indefinite causal structure is the quantum SWITCH ~\citep{PhysRevA.88.022318}, in which an auxiliary degree of freedom controls coherently the order in which operations act on system. This method enables higher order quantum operations that cannot be represented exactly using a standard quantum circuit \citep{PhysRevA.88.022318}. In addition, the quantum SWITCH has been shown to provide advantages in quantum computation~\citep{Colnaghi2012,PhysRevLett.113.250402}, communication~\citep{PhysRevLett.120.120502,2018arXiv180906655S,chiribella2018indefinite,2019arXiv190201807P}, metrology~\citep{2018arXiv181207508M,Zhao:19}, and, more recently, quantum refrigeration~\citep{PhysRevLett.125.070603}. The demonstration of the use of the quantum SWITCH, and thus ICO, has been claimed in recent experiments~\citep{Procopio2015,PhysRevLett.121.090503,goswami2018communicating,PhysRevLett.122.120504,PhysRevLett.124.030502}.


In this work, we obtain ICOs from the natural uncertainty of the moment at which two events $A$ and $B$ occur, e.g., the detection of spontaneous decays or the arrival of particles in an apparatus. We couple this uncertainty in a precise way so that the quantum SWITCH of operations ${\hat U}_A$ and ${\hat U}_B$ is implemented on a system $\cal S$. Thus, without resorting to unknown high-energy quantum gravitional effects, we provide a scheme in which ${\hat U}_A$ and ${\hat U}_B$ act in a temporal superposition that encompasses both possibilities ``${\hat U}_A$ acting before ${\hat U}_B$'' and ``${\hat U}_B$ acting before ${\hat U}_A$.'' These unitary operations can also be thought of as extensions of quantum channels in an ICO.

Although in this work we deal with the Schr\"odinger prescription of QM, where a time coordinate tracks the evolution of the system, we treat the instant of time when an event occurs as an observable. In this context, we will take advantage of certain aspects of the approach developed recently \citep{PhysRevA.103.012219}, where a space-time-symmetric formalism for events is proposed, to compute a quantum state with indefinite causal orders.



\paragraph*{Deriving ICO from superpositions in time---}
For the sake of clarity, we will show how indefinite causal orders can arise in a simple situation in which the decays of two uncorrelated atoms take place in a superposition of temporal orders. Consider the spontaneous emission of two atoms $A$ and $B$ with decay rates $\gamma_A$ and $\gamma_B$, respectively. The emitted photons from $A$ and $B$ trigger the unitary operations ${\hat U}_{A} \otimes \hat H$ and ${\hat U}_{B}\otimes {\hat \sigma}_z $ to act jointly on the system ($\cal S$) and the control qubit ($\cal C$), via a machine $\cal M$. Here, ${\hat H}$ and ${\hat \sigma}_z$ are the Hadamard and the Pauli-$z$ matrix in the basis $\{|0\rangle_{\cal C},|1\rangle_{\cal C}\}$, respectively. $\cal M$ registers the instant that the photons were received and performs their respective unitaries. To this end, the machine can have (among other subsystems) internal timers, such as the Salecker-Wigner-Peres-like timer $\cal T$~\cite{Pe,PhysRevA.103.012219}, in which an ideal quantum clock stops running when $\cal M$  the decay of $A$ or $B$. We assume that the operations take place instantaneously and that the free evolution of the system $\cal S$ and control $\cal C$ is the identity.


The system starts in the state
\begin{eqnarray}\label{initial}
 |\psi(t_{(0)})\rangle=|e,e\rangle_{AB}~|0,0\rangle_{\cal M}~|+\rangle_{\cal C}~ |\varphi\rangle_{\cal S},
\end{eqnarray}
where $|e,e\rangle_{AB}$ are the excited states of $A$ and $B$, and $|0,0\rangle_{\cal M}$ is the ready state of $\cal M$'s degrees of freedom that register the two different instants at which the operations take place. $|\varphi\rangle_{\cal S}$ is the initial state of $\cal S$ and  $|+\rangle_{\cal C}=(|0\rangle_{\cal C}+|1\rangle_{\cal C})/\sqrt{2}$ is chosen as the ready state of the control qubit. To prevent a dense notation, we avoid using the symbol $\otimes$. Let us calculate $|\psi(t)\rangle$ by breaking up the Schr\"odinger evolution into infinitesimal steps $\delta t=t_{(k+1)}-t_{(k)}\ll 1/\gamma_j$ ($k=1,2,\hdots$) in such a way that after the first step of the evolution, the initial state~(\ref{initial}) has evolved to
\begin{eqnarray}
\label{step1}
&|\psi&(t_{(1)})\rangle\nonumber\\
&=&\sqrt{1-\delta p_A}\sqrt{1-\delta p_B}~|e,e\rangle_{AB}~|0,0\rangle_{\cal M}~|+\rangle_{\cal C}~|\varphi\rangle_{\cal S}\nonumber\\
&+&\sqrt{\delta p_A}\sqrt{1-\delta p_B}~|g,e\rangle_{AB}~|t_{(1)},0\rangle_{\cal M}~|0\rangle_{\cal C}~{\hat U}_A~|\varphi\rangle_{\cal S}\nonumber\\
&+&\sqrt{1-\delta p_A}\sqrt{\delta p_B}~|e,g\rangle_{AB}~|0,t_{(1)}\rangle_{\cal M}~|1\rangle_{\cal C}~{\hat U}_B~|\varphi\rangle_{\cal S},\nonumber\\
\end{eqnarray}
with $\delta p_{i}=\gamma_{i}\delta t \ll 1$ being the probability of detecting the emission of $A/B$ in the interval $\delta t$. This ensures that the time step used is short enough such that we can neglect the probability $\delta p_A \delta p_B$ of both decays $A$ and $B$ happening simultaneously. 

In the first line of the expression for $|\psi(t_{(1)})\rangle$, we verify that with a high probability $(1-\delta p_{A})(1-\delta p_{B})$, $A$ and $B$ do not take decay (${\hat U}_A$ and ${\hat U}_B$ are not triggered), and hence the machine and the control qubit remain in their initial states. In the next lines, with probability $\delta p_{A}(1-\delta p_{B})$ [$\delta p_{B}(1-\delta p_{A})]$, $A$ [$B$] decays, and thus ${\hat U}_A\otimes {\hat H}$ [${\hat U}_B\otimes {\hat \sigma}_z$] acts on $\cal S\otimes\cal C$. The machine registers $t_{(1)}$ in different degrees of freedom, depending which operation is applied.

Calculating this evolution up to the $N$-th step, we have
\begin{widetext}
\begin{eqnarray}\label{stepN}
&|\psi&(t_{(N)})\rangle=\Gamma_A(t_{(N)},t_0)~\Gamma_B(t_{(N)},t_0)~|e,e\rangle_{AB}~|0,0\rangle_{\cal M}~|+\rangle_{\cal C}~|\varphi\rangle_{ \cal S}\nonumber\\
&+&\Gamma_B(t_{(N)},t_0)~\sum_{k=1}^{N} ~\tilde \chi_A(t_{(k)},t_{(k-1)})~|g,e\rangle_{AB}~|t_{(k)},0\rangle_{\cal M}~|0\rangle_{\cal C}~{\hat U}_{A}~|\varphi\rangle_{\cal S}\nonumber\\
&+&\Gamma_A(t_{(N)},t_0)~\sum_{k=1}^{N} ~\tilde \chi_B(t_{(k)},t_{(k-1)})~|e,g\rangle_{AB}~|0,t_{(k)}\rangle_{\cal M}~|1\rangle_{\cal C}~{\hat U}_B~|\varphi\rangle_{\cal S}\nonumber\\
&+&\sum_{\ell=k+1}^{N}\sum_{k=1}^{N-1} \tilde \chi_B(t_{(\ell)},t_{(\ell-1)})~\tilde \chi_A(t_{(k)},t_{(k-1)})~|g,g\rangle_{AB}~|t_{(k)},t_{(\ell)}\rangle_{\cal M}~|0\rangle_{\cal C}~{\hat U}_B~{\hat U}_A~|\varphi\rangle_{\cal S}\nonumber\\
&+&\sum_{\ell=k+1}^{N}\sum_{k=1}^{N-1} \tilde \chi_A(t_{(\ell)},t_{(\ell-1)})~\tilde \chi_B(t_{(k)},t_{(k-1)})~|g,g\rangle_{AB}~|t_{(\ell)},t_{(k)}\rangle_{\cal M}~|1\rangle_{\cal C}~{\hat U}_A~{\hat U}_B~|\varphi\rangle_{\cal S},
\end{eqnarray}
\end{widetext}
where $\Gamma_j(t_{(N)},t_0)=(1-\delta p_{j})^{N/2}$ is the probability amplitude for the the atoms $A/B$ not to decay in the interval $[t_0,t_{(N)}]$, and $\label{chi1}
\tilde \chi_j(t_{(k)},t_{(k-1)})=\sqrt{\delta p_{j}}~(1-\delta p_{j})^{(k-1)/2}$
is the probability amplitude for $A/B$ to decay in the interval $[t_{(k-1)},t_{(k)}]$. Here, we assume that $\delta p_j$ is the same for all intervals, but the generalization when this is not the case is straightforward.

In the first line of Eq.~(\ref{stepN}), we verify that $A$ and $B$ do not decay in the interval $(t_0,t_{(N)}]$, thus $\cal M$ remains in the initial state. In contrast, in the second (third) line, $A$ ($B$) decays in the interval $[t_{(k-1)},t_{(k)}]$, $\cal M$ registers $t_{(k)}$ in its first (second) degree of freedom, and $B$ ($A$) remains in the excited state. Finally, $A$ and $B$ are in the ground state in the fourth (fifth) line, with $A$ ($B$) decaying at the instant $t_{(k)}$, and $B$ ($A$) jumping later at the instant $t_{(\ell)}$, and $\cal M$ registering $t_{(k)}$ and $t_{(\ell)}$.

We are interested in $|\psi(t)\rangle$ from the moment we can guarantee that both detections $A$ and $B$ have indeed occurred, and hence both ${\hat U}_A$ and ${\hat U}_B$ have acted on $\cal S$. We therefore evaluate $|\psi(t_{(N)})\rangle$ when $t_{(N)} \gg t_0 +{\rm max}(1/\gamma_A,1/\gamma_B)$, so that $\Gamma_j(t_{(N)},t_0)\approx 0$. For simplicity, let us take $t_{(N)}\rightarrow \infty$. In this scenario, as the atoms $A$ and $B$ are no longer correlated to the other systems, the state of $\cal MSC$ [$|\psi\rangle\equiv|\psi(t_{(N)}\rightarrow \infty)\rangle$] in the continuous limit becomes
\begin{eqnarray}\label{eq:final}
|\psi\rangle_{\cal MSC}=\int_{t_0}^\infty&&\int_{t_0}^\infty~dt_A dt_B~\chi_{AB}(t_A,t_B)~|t_A,t_B\rangle_{\cal M}~\nonumber\\
\otimes\bigg[&&\Theta(t_B-t_A)~|0\rangle_{\cal C}~{\hat U}_B~{\hat U}_A~|\varphi\rangle_{\cal S}\nonumber\\
+&&\Theta(t_A-t_B)~|1\rangle_{\cal C}~{\hat U}_A~{\hat U}_B~|\varphi\rangle_{\cal S}\bigg],
\end{eqnarray}
where $\chi_{AB}(t_A,t_B)=\chi_A(t_A)\chi_B(t_B)$ is the continuous version of $\tilde\chi$, and $\Theta(t_x-t_y)=1$ for $t_x>t_y$, and $0$ otherwise. Here, $\chi_j(t_j)$ is the probability density amplitude for the detection $j$ to happen at the instant $t_j$, which represents the continuous limit of Eq.~(\ref{chi1}). In Eq.~(\ref{eq:final}), besides the indefinite causal order between ${\hat U}_A$ and ${\hat U}_B$, $|\psi\rangle_{\cal MSC}$ depicts a superposition of all possibles instants $t_A$ and $t_B$ [with weight $\chi_{AB}(t_A,t_B)$] at which ${\hat U}_A$ and ${\hat U}_B$ can act on $\cal S$.

To obtain $\chi_j(t_j)$, notice that the probability for the atoms to decay in the interval $(t_{(k)},t_{(k+1)}]$ can be calculated either by $\chi_j(t_{(k)},t_{(k+1)})$ squared [Eq.~(\ref{chi1})] or by the modulus square of $\chi_j(t_j)$ at the instant $t_{(k)}$ multiplied by $\delta t$. By equating these two expressions, as $\delta p_j \ll 1$, we have
$\delta p_{j}(k-1)\delta p_{j}\approx |\chi_{j}(t_{(k)})|^2~\delta t.$
Then, since $\delta p_j$ is time-independent, let us take $\delta p_j=\gamma_j \delta t$. By considering $t_{(k)}-t_0=k\delta t$ and isolating $|\chi_{j}(t_{(k)})|^2$ in its expression above, we obtain that the probability density amplitude for an atom to decay at the instant $t_j$ can be written as
$\chi_j(t_j)=\sqrt{\gamma_j} ~{\exp }{-\gamma_j (t_j-t_0)/2}$
for $t_j\geq t_0$, and $\chi_j(t_j)=0$ otherwise.

It is worth noticing that since we chose a time-independent ${\delta p}_j$, $\chi_j(t_j)$ is simply the time derivative of the traditional spontaneous decay amplitude. Also, one can verify that for more complex events --- where, for instance, the events associated with $A$ and $B$ are correlated --- $|\psi\rangle_{\cal MSC}$ has the same format as Eq.~(\ref{eq:final}) but with a non-separable time probability amplitude $\chi_{AB}(t_A,t_B)$.


It is evident that the superposition in Eq.~(\ref{eq:final}) contains branches in which decay $A$ happens first and ${\hat U}_A$ acts first and branches in which decay $B$ happens first and ${\hat U}_B$ acts first. The causal order in this situation is thus indefinite. However, we have not yet recovered the mathematical form of the quantum SWITCH, because the final state of the system $\cal S$ and control $\cal C$ are still entangled with the machine $\cal M$, and depend on the distribution $\chi_{AB}$. Nevertheless, by including $\cal C$ and $\cal M$ as part of a composite subsystem $\cal {M'}$, and defining the states
\begin{eqnarray}\label{state0}
\ket{0}_{\cal M'}=\sqrt{2}\int_{t_0}^\infty dt_A\int_{t_A}^\infty dt_B~\chi_{AB}(t_A,t_B)~|t_A,t_B\rangle_{\cal M}|0\rangle_{\cal C}\nonumber\\
\end{eqnarray}
and
\begin{eqnarray}\label{state1}
\ket{1}_{\cal M'}=\sqrt{2}\int_{t_0}^\infty dt_B\int_{t_B}^\infty dt_A~ \chi_{AB}(t_A,t_B)~|t_A,t_B\rangle_{\cal M}|1\rangle_{\cal C}, \nonumber\\
\end{eqnarray}
Eq.~(\ref{eq:final}) acquires, in principle, the mathematical structure of the quantum SWITCH,
\begin{align}\label{eq:ICO1}
\ket{\psi}_{{\cal M'}{\cal S}}=\frac{1}{\sqrt{2}}\Big(\ket{0}_{\cal M'}~\hat{U}_B \hat{U}_A~  + \ket{1}_{\cal M'}~{\hat{U}}_A {\hat{U}}_B ~ \Big)\ket{\varphi}_{\cal S},
\end{align}
with $_{\cal M'}\langle 0|1 \rangle_{\cal M'}=0$ and $_{\cal M'}\langle 0|0 \rangle_{\cal M'}=_{\cal M'}\langle 1|1 \rangle_{\cal M'}=0$. Note that for $|\psi\rangle_{{\cal M'}{\cal S}}$ to be useful, one should be able to distinguish between the states $|0\rangle_{\cal M'}$ and $|1\rangle_{\cal M'}$. On the other hand,  if one wants to have the traditional quantum SWITCH, with a simpler control system (not involving the coherence of many levels), one should, for instance, disentangle $\cal M$ and $\cal C$ (which is the second strategy we will propose next). We proceed to consider two physical realizations which are engineered such that the original quantum SWITCH is recovered.

\paragraph*{Realization 1: time-bin entangled photons---}


\begin{figure}[ht]
 \includegraphics[width=81mm]{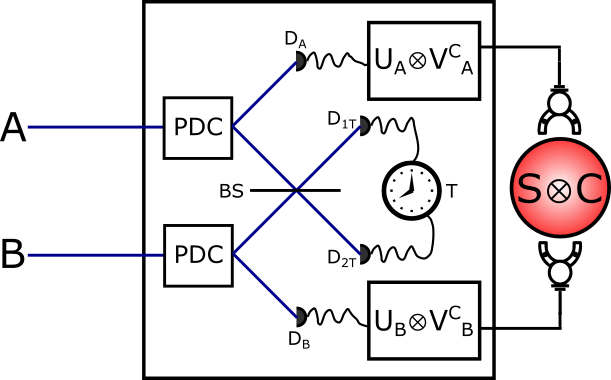}
 \caption{Photons from the decays of atoms $A$ and $B$ travel to the machine, which operates on the system and control qubit with the corresponding unitary when a photon is received. A photon enters the machine and undergoes parametric down conversion (PDC), being split into two photons. One of these photons travels to a detector (${\cal D}_A / {\cal D}_{B}$) which triggers the unitary associated with that particular atom on the system ($\hat{U}_{A/B}$ on $\cal S$) and the control ($\hat{V}^\text{C}_{A/B}$ on $\cal C$). The other travels, via a beam splitter (BS), to one of two detectors (${\cal D}_{1{\cal T}}$ or ${\cal D}_{2{\cal T}}$), which cause the timer ($\cal T$) to record the time of arrival of the photon, but not which photon arrived first.}
 \label{fig:OverallFig}
\end{figure}

\begin{figure*}[ht]
 \includegraphics[width=160mm]{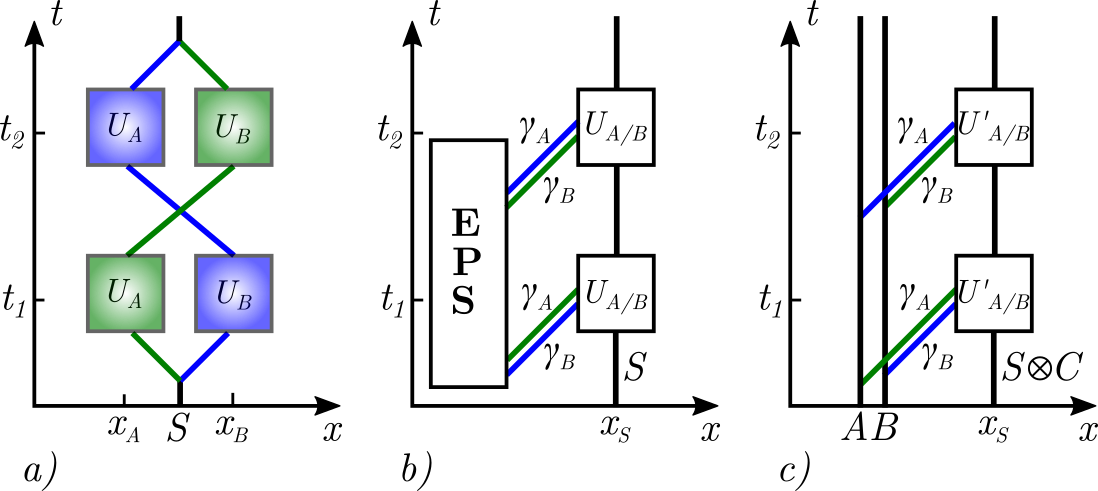}
 \caption{Spacetime diagrams for three implementations of the quantum SWITCH a) An implementation using a superposition of spatial trajectories to achieve the SWITCH must have 4 spacetime points associated with the operations. The spatial degree of freedom of the system is in a superposition (green and blue) and so the operations all take place at 4 distinct spacetime points: [$(t_1,x_A)$,$(t_1,x_B)$,$(t_2,x_A)$,$(t_2,x_B)$]. b) and c) show our scheme using time bin entangled photons (EPS) (b) and decaying atoms $A$ and $B$ (c). An implementation using our scheme relies upon superpositions in time, rather that in space, of the events. It is therefore possible to construct a SWITCH in which there are only two spacetime points associated with the operations.}
 \label{diag2}
\end{figure*}

For the first strategy, we consider a timer state that describes time-bin entangled photons, such as the ones produced in \citep{timebinnature,timebin}: $\ket{\varphi}_{AB}=\frac{1}{\sqrt{2}} \left(\ket{e}_A\ket{l}_B + \ket{l}_A\ket{e}_B\right)$. Such states can be considered superposed in time with the labels $e$ and $l$ denoting early (forward photon) and late (backward photon). Considering time-bin entangled photons ($A$ and $B$), which trigger the operations ${\hat U}_A$ and ${\hat U}_B$ via the machine $\cal M$ as previously discussed, the time probability amplitude of Eqs.~(\ref{state0}) and (\ref{state1}) is then given by $|\chi_{AB}(t_A,t_B)|^2 = \frac{1}{2}\big[\delta(t_A-t_e)\delta(t_B-t_l) + \delta(t_A-t_l)\delta(t_B-t_e)\big]$, where $t_e$ and $t_l$ are the early and late times respectively. Then, there is a superposition of operation $A$ happening at time $t_e$ then operation $B$ at time $t_l$, and of operation $B$ happening at time $t_e$ then operation $A$ happening at time $t_l$. Performing the integration from Eqs.~(\ref{state0}) and (\ref{state1}), we obtain
\begin{eqnarray}\label{timebin}
|0\rangle_{\cal M'}=|t_e,t_l\rangle_{\cal M}~|0 \rangle_{\cal C}~~{\rm and}~~|1\rangle_{\cal M'}=|t_l,t_e\rangle_{\cal M}~|1\rangle_{\cal C}.~~~~~~
\end{eqnarray}
Notice that in the particular case of this first strategy, we can neglect $\cal C$, since $\cal M$ can play the role of the control qubit. By substituting theses states into Eq.~(\ref{eq:ICO1}), we have completely recovered the action of the quantum SWITCH of the gates on an initial state $\ket{\varphi}_{\cal S}$. Later, we will argue that this scheme constitutes a true implementation of the quantum SWITCH supermap.

\paragraph*{Realization 2: decaying atoms---}

Although we have already proposed a method using time-bin entangled photons, it might be objected that the control over this set of states is too difficult to be practical. Also, it may appear that the scheme depends entirely on time-bin entanglement as a resource. To mitigate these concerns, we show here that one can realise the usual quantum SWITCH using the decays of two unentangled atoms, as previously described. To this end, we have to decouple $\cal M$ from the rest. First, if the two atoms are positioned so that the emitted photons enter the machine at different positions, then the machine can record the order in which the two decays took place. The diagram for the scheme is presented in Fig.~\ref{fig:OverallFig}, with $\hat{V}^C_{A}={\hat H}$ and $\hat{V}^C_{B}={\hat \sigma}_z$.

By inspecting Eqs.~(\ref{state0}) and (\ref{state1}) [or Eq.~(\ref{eq:final})] for $\chi_{AB}(t_A,t_B)=\chi_A(t_A)\chi_B(t_B)$ with $\chi_A(x)=\chi_B(x)$, we verify that in order to decouple the machine states $|t_A,t_B\rangle_{\cal M}$ from the system and the control qubit, it is sufficient that $|t_A,t_B\rangle_{\cal M}$ = $|t_B,t_A\rangle_{\cal M}$. This is the reason for the beam splitter before the timer in Fig.~\ref{fig:OverallFig}. The timer records the time of arrival of both photons, but not which one arrived first, because the beam splitter sends both photons to the two detectors ${\cal D}_{1{\cal T}}$ and ${\cal D}_{2{\cal T}}$ with equal probability.

As $|t_A,t_B\rangle_{\cal M}$ = $|t_B,t_A\rangle_{\cal M}$, we can perform the change of variables $t_A \rightleftharpoons t_B$ in the integral of the state $|1\rangle_{\cal M'}$ of Eq.~(\ref{state1}) [or, similarly, in the second integral of Eq.~(\ref{eq:final})] to obtain $|\psi\rangle_{{\cal M'}{\cal S}}=|\psi\rangle_{{\cal M}{\cal C}{\cal S}}$ written as 
\begin{eqnarray}\label{eq:TwoParts2}
|\psi\rangle_{\cal MSC}=\int_{t_0}^\infty dt_A\int_{t_A}^\infty dt_B~\sqrt{2}~\chi(t_A)\chi(t_B)~|t_A,t_B\rangle_{\cal M}~\nonumber\\
\otimes~\frac{1}{\sqrt{2}}\bigg[|0\rangle_{\cal C}~{\hat U}_B~{\hat U}_A~|\varphi\rangle_{\cal S} + |1\rangle_{\cal C}~{\hat U}_A~{\hat U}_B~|\varphi\rangle_{\cal S} \bigg].~~~~~
\end{eqnarray}
Now the machine ($\cal M$) and the rest of the system ($\cal {SC}$) are in a product state. We are thus free to discard $\cal M$. What remains is the desired result - the expression for the quantum SWITCH with a simple two-level control system.

\paragraph*{Discussion---}

This work was motivated in part by debate about what constitutes a true implementation of an indefinite causal order, and whether one can be achieved without superpositions of space-time in a quantum gravitational framework. We will now ask whether our quantum SWITCH protocol represents a true ICO and in what respects it is similar and different to other implementations. 

A similarity between our scheme and the quantum SWITCHes that have already been implemented in experiments is that they share the same representation in the process matrix formalism. This formalism gives a framework for quantum mechanics in situations where the causal structure is indefinite. The quantum SWITCH has been studied \citep{Oreshkov2019timedelocalized} using this formalism, including cases where the operations are performed on time-delocalized subsystems. The process matrix specifies the input and output Hilbert spaces on which the quantum operations taking place during the process act. In our scheme, the nature of the time-delocalization is explicit - the input and output Hilbert spaces are given by the Hilbert spaces associated with each time slice of Fig. \ref{diag2}. It can be seen therefore that the process described above is represented by the same process matrix as discussed in \citep{Oreshkov2019timedelocalized}, implying that this scheme is a true implementation of the quantum SWITCH.

We now comment on the major difference between our scheme and previously implemented protocols. It has been argued \citep{Paunkovic2020causalorders} that the photonic implementations of the quantum SWITCH are disqualified from being genuine ICOs because they involve four spacetime events. It is therefore important to address whether our scheme can be criticized in the same way.  The two situations are represented in Fig. \ref{diag2}. Diagram a) represents schematically implementations that rely on superposition in space, such as the photonic implementations mentioned above. Diagrams b) and c) represent two versions of out protocol, relying on superpositions in time.

From these diagrams, we can see that in Ref. \citep{PhysRevLett.121.090503}, for example, the control
qubit is entangled with the spatial degree of freedom of the system. Then, in one
branch Alice applies the $\hat{U}_A$ controlled unitary at spacetime event $(t_1,x_A)$ and Bob applies the controlled unitary $\hat{U}_B$ at  $(t_2,x_B)$, while in the other branch Bob acts at $(t_1,x_B)$ and Alice at $(t_2,x_A)$. Then the system and control are recombined. We therefore see that four spacetime events are relevant to describe the process that the system undergoes.

In our approach, however, the system is not required to become delocalised
in space. The unitaries are performed at the same spatial location. It is not
difficult to identify then that there are only two space-time points where the system
undergoes a transformation. These are $(t_1,x_S)$ and $(t_2,x_S)$. It has been claimed \citep{Paunkovic2020causalorders} that this difference in the number of meaningful spacetime points is important for determining whether a true ICO has been achieved. We therefore believe that this protocol really does implement a genuine quantum SWITCH. 

This work only considers how the quantum SWITCH might arise as a result of superpositions in time. However, the SWITCH is not the only indefinite causal structure. Further work is needed to discover which other structures can arise in a similar way, and whether there are any allowed causal structures which cannot be realised using superpositions in time.

Since processes that are indefinite in time occur widely in nature, and we have shown how ICO can arise fairly simply when such superpositions exist, this work suggests that exotic causal structures might be more widespread than previously thought. This encourages the search for exotic causal structures arising naturally. It is possible that certain natural phenomena might be able to be explained only with reference to such causal structures.

\paragraph*{Acknowledgements---}
We would like to thank Aditya Iyer and Sam Kuypers for helpful discussions and comments. DF is supported by the EPSRC (UK) and by M squared. VV thanks the National Research Foundation, Prime Minister’s Office, Singapore, under its Competitive Research Programme (CRP Award No. NRF- CRP14-2014-02) and administered by Centre for Quantum Technologies, National University of Singapore. EOD acknowledges financial support from Conselho Nacional de Desenvolvimento Científico e Tecnológico (CNPq) through its program  09/2020 (Grant No. 315759/2020-8) and Coordenação de Aperfeiçoamento de Pessoal de Nível Superior (CAPES) through its program GPCT - 17/2016 (Grant No. 88887.312745/2018-00).

\bibliography{references}

\begin{thebibliography}{24}%
\makeatletter
\providecommand \@ifxundefined [1]{%
 \@ifx{#1\undefined}
}%
\providecommand \@ifnum [1]{%
 \ifnum #1\expandafter \@firstoftwo
 \else \expandafter \@secondoftwo
 \fi
}%
\providecommand \@ifx [1]{%
 \ifx #1\expandafter \@firstoftwo
 \else \expandafter \@secondoftwo
 \fi
}%
\providecommand \natexlab [1]{#1}%
\providecommand \enquote  [1]{``#1''}%
\providecommand \bibnamefont  [1]{#1}%
\providecommand \bibfnamefont [1]{#1}%
\providecommand \citenamefont [1]{#1}%
\providecommand \href@noop [0]{\@secondoftwo}%
\providecommand \href [0]{\begingroup \@sanitize@url \@href}%
\providecommand \@href[1]{\@@startlink{#1}\@@href}%
\providecommand \@@href[1]{\endgroup#1\@@endlink}%
\providecommand \@sanitize@url [0]{\catcode `\\12\catcode `\$12\catcode
  `\&12\catcode `\#12\catcode `\^12\catcode `\_12\catcode `\%12\relax}%
\providecommand \@@startlink[1]{}%
\providecommand \@@endlink[0]{}%
\providecommand \url  [0]{\begingroup\@sanitize@url \@url }%
\providecommand \@url [1]{\endgroup\@href {#1}{\urlprefix }}%
\providecommand \urlprefix  [0]{URL }%
\providecommand \Eprint [0]{\href }%
\providecommand \doibase [0]{http://dx.doi.org/}%
\providecommand \selectlanguage [0]{\@gobble}%
\providecommand \bibinfo  [0]{\@secondoftwo}%
\providecommand \bibfield  [0]{\@secondoftwo}%
\providecommand \translation [1]{[#1]}%
\providecommand \BibitemOpen [0]{}%
\providecommand \bibitemStop [0]{}%
\providecommand \bibitemNoStop [0]{.\EOS\space}%
\providecommand \EOS [0]{\spacefactor3000\relax}%
\providecommand \BibitemShut  [1]{\csname bibitem#1\endcsname}%
\let\auto@bib@innerbib\@empty
\bibitem [{\citenamefont {Hardy}(2007)}]{Hardy_2007}%
  \BibitemOpen
  \bibfield  {author} {\bibinfo {author} {\bibfnamefont {L.}~\bibnamefont
  {Hardy}},\ }\href {\doibase 10.1088/1751-8113/40/12/s12} {\bibfield
  {journal} {\bibinfo  {journal} {Journal of Physics A: Mathematical and
  Theoretical}\ }\textbf {\bibinfo {volume} {40}},\ \bibinfo {pages} {3081}
  (\bibinfo {year} {2007})}\BibitemShut {NoStop}%
\bibitem [{\citenamefont {Oreshkov}\ \emph {et~al.}(2012)\citenamefont
  {Oreshkov}, \citenamefont {Costa},\ and\ \citenamefont {Brukner}}]{Ore}%
  \BibitemOpen
  \bibfield  {author} {\bibinfo {author} {\bibfnamefont {O.}~\bibnamefont
  {Oreshkov}}, \bibinfo {author} {\bibfnamefont {F.}~\bibnamefont {Costa}}, \
  and\ \bibinfo {author} {\bibfnamefont {C.}~\bibnamefont {Brukner}},\ }\href
  {\doibase 10.1038/ncomms2076} {\bibfield  {journal} {\bibinfo  {journal}
  {Nat. Comms.}\ }\textbf {\bibinfo {volume} {3}},\ \bibinfo {pages} {1092}
  (\bibinfo {year} {2012})}\BibitemShut {NoStop}%
\bibitem [{\citenamefont {Giacomini}\ \emph {et~al.}(2016)\citenamefont
  {Giacomini}, \citenamefont {Castro-Ruiz},\ and\ \citenamefont
  {Brukner}}]{Giacomini_2016}%
  \BibitemOpen
  \bibfield  {author} {\bibinfo {author} {\bibfnamefont {F.}~\bibnamefont
  {Giacomini}}, \bibinfo {author} {\bibfnamefont {E.}~\bibnamefont
  {Castro-Ruiz}}, \ and\ \bibinfo {author} {\bibfnamefont {C.}~\bibnamefont
  {Brukner}},\ }\href {\doibase 10.1088/1367-2630/18/11/113026} {\bibfield
  {journal} {\bibinfo  {journal} {New Journal of Physics}\ }\textbf {\bibinfo
  {volume} {18}},\ \bibinfo {pages} {113026} (\bibinfo {year}
  {2016})}\BibitemShut {NoStop}%
\bibitem [{\citenamefont {Chiribella}\ \emph {et~al.}(2013)\citenamefont
  {Chiribella}, \citenamefont {D'Ariano}, \citenamefont {Perinotti},\ and\
  \citenamefont {Valiron}}]{PhysRevA.88.022318}%
  \BibitemOpen
  \bibfield  {author} {\bibinfo {author} {\bibfnamefont {G.}~\bibnamefont
  {Chiribella}}, \bibinfo {author} {\bibfnamefont {G.~M.}\ \bibnamefont
  {D'Ariano}}, \bibinfo {author} {\bibfnamefont {P.}~\bibnamefont {Perinotti}},
  \ and\ \bibinfo {author} {\bibfnamefont {B.}~\bibnamefont {Valiron}},\ }\href
  {\doibase 10.1103/PhysRevA.88.022318} {\bibfield  {journal} {\bibinfo
  {journal} {Phys. Rev. A}\ }\textbf {\bibinfo {volume} {88}},\ \bibinfo
  {pages} {022318} (\bibinfo {year} {2013})}\BibitemShut {NoStop}%
\bibitem [{\citenamefont {Colnaghi}\ \emph {et~al.}(2012)\citenamefont
  {Colnaghi}, \citenamefont {D'Ariano}, \citenamefont {Facchini},\ and\
  \citenamefont {Perinotti}}]{Colnaghi2012}%
  \BibitemOpen
  \bibfield  {author} {\bibinfo {author} {\bibfnamefont {T.}~\bibnamefont
  {Colnaghi}}, \bibinfo {author} {\bibfnamefont {G.~M.}\ \bibnamefont
  {D'Ariano}}, \bibinfo {author} {\bibfnamefont {S.}~\bibnamefont {Facchini}},
  \ and\ \bibinfo {author} {\bibfnamefont {P.}~\bibnamefont {Perinotti}},\
  }\href {\doibase 10.1016/j.physleta.2012.08.028} {\bibfield  {journal}
  {\bibinfo  {journal} {Physics Letters A}\ }\textbf {\bibinfo {volume}
  {376}},\ \bibinfo {pages} {2940} (\bibinfo {year} {2012})}\BibitemShut
  {NoStop}%
\bibitem [{\citenamefont {Ara\'ujo}\ \emph {et~al.}(2014)\citenamefont
  {Ara\'ujo}, \citenamefont {Costa},\ and\ \citenamefont
  {Brukner}}]{PhysRevLett.113.250402}%
  \BibitemOpen
  \bibfield  {author} {\bibinfo {author} {\bibfnamefont {M.}~\bibnamefont
  {Ara\'ujo}}, \bibinfo {author} {\bibfnamefont {F.}~\bibnamefont {Costa}}, \
  and\ \bibinfo {author} {\bibfnamefont {i.~c.~v.}\ \bibnamefont {Brukner}},\
  }\href {\doibase 10.1103/PhysRevLett.113.250402} {\bibfield  {journal}
  {\bibinfo  {journal} {Phys. Rev. Lett.}\ }\textbf {\bibinfo {volume} {113}},\
  \bibinfo {pages} {250402} (\bibinfo {year} {2014})}\BibitemShut {NoStop}%
\bibitem [{\citenamefont {Ebler}\ \emph {et~al.}(2018)\citenamefont {Ebler},
  \citenamefont {Salek},\ and\ \citenamefont
  {Chiribella}}]{PhysRevLett.120.120502}%
  \BibitemOpen
  \bibfield  {author} {\bibinfo {author} {\bibfnamefont {D.}~\bibnamefont
  {Ebler}}, \bibinfo {author} {\bibfnamefont {S.}~\bibnamefont {Salek}}, \ and\
  \bibinfo {author} {\bibfnamefont {G.}~\bibnamefont {Chiribella}},\ }\href
  {\doibase 10.1103/PhysRevLett.120.120502} {\bibfield  {journal} {\bibinfo
  {journal} {Phys. Rev. Lett.}\ }\textbf {\bibinfo {volume} {120}},\ \bibinfo
  {pages} {120502} (\bibinfo {year} {2018})}\BibitemShut {NoStop}%
\bibitem [{\citenamefont {{Salek}}\ \emph {et~al.}(2018)\citenamefont
  {{Salek}}, \citenamefont {{Ebler}},\ and\ \citenamefont
  {{Chiribella}}}]{2018arXiv180906655S}%
  \BibitemOpen
  \bibfield  {author} {\bibinfo {author} {\bibfnamefont {S.}~\bibnamefont
  {{Salek}}}, \bibinfo {author} {\bibfnamefont {D.}~\bibnamefont {{Ebler}}}, \
  and\ \bibinfo {author} {\bibfnamefont {G.}~\bibnamefont {{Chiribella}}},\
  }\href@noop {} {\bibfield  {journal} {\bibinfo  {journal} {arXiv e-prints}\
  ,\ \bibinfo {eid} {arXiv:1809.06655}} (\bibinfo {year} {2018})},\ \Eprint
  {http://arxiv.org/abs/1809.06655} {arXiv:1809.06655 [quant-ph]} \BibitemShut
  {NoStop}%
\bibitem [{\citenamefont {Chiribella}\ \emph {et~al.}(2018)\citenamefont
  {Chiribella}, \citenamefont {Banik}, \citenamefont {Bhattacharya},
  \citenamefont {Guha}, \citenamefont {Alimuddin}, \citenamefont {Roy},
  \citenamefont {Saha}, \citenamefont {Agrawal},\ and\ \citenamefont
  {Kar}}]{chiribella2018indefinite}%
  \BibitemOpen
  \bibfield  {author} {\bibinfo {author} {\bibfnamefont {G.}~\bibnamefont
  {Chiribella}}, \bibinfo {author} {\bibfnamefont {M.}~\bibnamefont {Banik}},
  \bibinfo {author} {\bibfnamefont {S.~S.}\ \bibnamefont {Bhattacharya}},
  \bibinfo {author} {\bibfnamefont {T.}~\bibnamefont {Guha}}, \bibinfo {author}
  {\bibfnamefont {M.}~\bibnamefont {Alimuddin}}, \bibinfo {author}
  {\bibfnamefont {A.}~\bibnamefont {Roy}}, \bibinfo {author} {\bibfnamefont
  {S.}~\bibnamefont {Saha}}, \bibinfo {author} {\bibfnamefont {S.}~\bibnamefont
  {Agrawal}}, \ and\ \bibinfo {author} {\bibfnamefont {G.}~\bibnamefont
  {Kar}},\ }\href@noop {} {\enquote {\bibinfo {title} {Indefinite causal order
  enables perfect quantum communication with zero capacity channel},}\ }
  (\bibinfo {year} {2018}),\ \Eprint {http://arxiv.org/abs/1810.10457}
  {arXiv:1810.10457 [quant-ph]} \BibitemShut {NoStop}%
\bibitem [{\citenamefont {{Procopio}}\ \emph {et~al.}(2019)\citenamefont
  {{Procopio}}, \citenamefont {{Delgado}}, \citenamefont {{Enriquez}},
  \citenamefont {{Belabas}},\ and\ \citenamefont
  {{Levenson}}}]{2019arXiv190201807P}%
  \BibitemOpen
  \bibfield  {author} {\bibinfo {author} {\bibfnamefont {L.~M.}\ \bibnamefont
  {{Procopio}}}, \bibinfo {author} {\bibfnamefont {F.}~\bibnamefont
  {{Delgado}}}, \bibinfo {author} {\bibfnamefont {M.}~\bibnamefont
  {{Enriquez}}}, \bibinfo {author} {\bibfnamefont {N.}~\bibnamefont
  {{Belabas}}}, \ and\ \bibinfo {author} {\bibfnamefont {J.~A.}\ \bibnamefont
  {{Levenson}}},\ }\href@noop {} {\bibfield  {journal} {\bibinfo  {journal}
  {arXiv e-prints}\ ,\ \bibinfo {eid} {arXiv:1902.01807}} (\bibinfo {year}
  {2019})},\ \Eprint {http://arxiv.org/abs/1902.01807} {arXiv:1902.01807
  [quant-ph]} \BibitemShut {NoStop}%
\bibitem [{\citenamefont {{Mukhopadhyay}}\ \emph {et~al.}(2018)\citenamefont
  {{Mukhopadhyay}}, \citenamefont {{Gupta}},\ and\ \citenamefont
  {{Pati}}}]{2018arXiv181207508M}%
  \BibitemOpen
  \bibfield  {author} {\bibinfo {author} {\bibfnamefont {C.}~\bibnamefont
  {{Mukhopadhyay}}}, \bibinfo {author} {\bibfnamefont {M.~K.}\ \bibnamefont
  {{Gupta}}}, \ and\ \bibinfo {author} {\bibfnamefont {A.~K.}\ \bibnamefont
  {{Pati}}},\ }\href@noop {} {\bibfield  {journal} {\bibinfo  {journal} {arXiv
  e-prints}\ ,\ \bibinfo {eid} {arXiv:1812.07508}} (\bibinfo {year} {2018})},\
  \Eprint {http://arxiv.org/abs/1812.07508} {arXiv:1812.07508 [quant-ph]}
  \BibitemShut {NoStop}%
\bibitem [{\citenamefont {Zhao}\ and\ \citenamefont {Giulio}(2019)}]{Zhao:19}%
  \BibitemOpen
  \bibfield  {author} {\bibinfo {author} {\bibfnamefont {X.}~\bibnamefont
  {Zhao}}\ and\ \bibinfo {author} {\bibfnamefont {C.}~\bibnamefont {Giulio}},\
  }in\ \href {\doibase 10.1364/QIM.2019.F5A.23} {\emph {\bibinfo {booktitle}
  {Quantum Information and Measurement (QIM) V: Quantum Technologies}}}\
  (\bibinfo  {publisher} {Optical Society of America},\ \bibinfo {year}
  {2019})\ p.\ \bibinfo {pages} {F5A.23}\BibitemShut {NoStop}%
\bibitem [{\citenamefont {Felce}\ and\ \citenamefont
  {Vedral}(2020)}]{PhysRevLett.125.070603}%
  \BibitemOpen
  \bibfield  {author} {\bibinfo {author} {\bibfnamefont {D.}~\bibnamefont
  {Felce}}\ and\ \bibinfo {author} {\bibfnamefont {V.}~\bibnamefont {Vedral}},\
  }\href {\doibase 10.1103/PhysRevLett.125.070603} {\bibfield  {journal}
  {\bibinfo  {journal} {Phys. Rev. Lett.}\ }\textbf {\bibinfo {volume} {125}},\
  \bibinfo {pages} {070603} (\bibinfo {year} {2020})}\BibitemShut {NoStop}%
\bibitem [{\citenamefont {Procopio}\ \emph {et~al.}(2015)\citenamefont
  {Procopio}, \citenamefont {Moqanaki}, \citenamefont {Ara{\'{u}}jo},
  \citenamefont {Costa}, \citenamefont {Calafell}, \citenamefont {Dowd},
  \citenamefont {Hamel}, \citenamefont {Rozema}, \citenamefont {Brukner},\ and\
  \citenamefont {Walther}}]{Procopio2015}%
  \BibitemOpen
  \bibfield  {author} {\bibinfo {author} {\bibfnamefont {L.~M.}\ \bibnamefont
  {Procopio}}, \bibinfo {author} {\bibfnamefont {A.}~\bibnamefont {Moqanaki}},
  \bibinfo {author} {\bibfnamefont {M.}~\bibnamefont {Ara{\'{u}}jo}}, \bibinfo
  {author} {\bibfnamefont {F.}~\bibnamefont {Costa}}, \bibinfo {author}
  {\bibfnamefont {I.~A.}\ \bibnamefont {Calafell}}, \bibinfo {author}
  {\bibfnamefont {E.~G.}\ \bibnamefont {Dowd}}, \bibinfo {author}
  {\bibfnamefont {D.~R.}\ \bibnamefont {Hamel}}, \bibinfo {author}
  {\bibfnamefont {L.~A.}\ \bibnamefont {Rozema}}, \bibinfo {author}
  {\bibfnamefont {{\v{C}}.}~\bibnamefont {Brukner}}, \ and\ \bibinfo {author}
  {\bibfnamefont {P.}~\bibnamefont {Walther}},\ }\href {\doibase
  10.1038/ncomms8913} {\bibfield  {journal} {\bibinfo  {journal} {Nature
  Communications}\ }\textbf {\bibinfo {volume} {6}} (\bibinfo {year} {2015}),\
  10.1038/ncomms8913}\BibitemShut {NoStop}%
\bibitem [{\citenamefont {Goswami}\ \emph
  {et~al.}(2018{\natexlab{a}})\citenamefont {Goswami}, \citenamefont
  {Giarmatzi}, \citenamefont {Kewming}, \citenamefont {Costa}, \citenamefont
  {Branciard}, \citenamefont {Romero},\ and\ \citenamefont
  {White}}]{PhysRevLett.121.090503}%
  \BibitemOpen
  \bibfield  {author} {\bibinfo {author} {\bibfnamefont {K.}~\bibnamefont
  {Goswami}}, \bibinfo {author} {\bibfnamefont {C.}~\bibnamefont {Giarmatzi}},
  \bibinfo {author} {\bibfnamefont {M.}~\bibnamefont {Kewming}}, \bibinfo
  {author} {\bibfnamefont {F.}~\bibnamefont {Costa}}, \bibinfo {author}
  {\bibfnamefont {C.}~\bibnamefont {Branciard}}, \bibinfo {author}
  {\bibfnamefont {J.}~\bibnamefont {Romero}}, \ and\ \bibinfo {author}
  {\bibfnamefont {A.~G.}\ \bibnamefont {White}},\ }\href {\doibase
  10.1103/PhysRevLett.121.090503} {\bibfield  {journal} {\bibinfo  {journal}
  {Phys. Rev. Lett.}\ }\textbf {\bibinfo {volume} {121}},\ \bibinfo {pages}
  {090503} (\bibinfo {year} {2018}{\natexlab{a}})}\BibitemShut {NoStop}%
\bibitem [{\citenamefont {Goswami}\ \emph
  {et~al.}(2018{\natexlab{b}})\citenamefont {Goswami}, \citenamefont {Cao},
  \citenamefont {Paz-Silva}, \citenamefont {Romero},\ and\ \citenamefont
  {White}}]{goswami2018communicating}%
  \BibitemOpen
  \bibfield  {author} {\bibinfo {author} {\bibfnamefont {K.}~\bibnamefont
  {Goswami}}, \bibinfo {author} {\bibfnamefont {Y.}~\bibnamefont {Cao}},
  \bibinfo {author} {\bibfnamefont {G.~A.}\ \bibnamefont {Paz-Silva}}, \bibinfo
  {author} {\bibfnamefont {J.}~\bibnamefont {Romero}}, \ and\ \bibinfo {author}
  {\bibfnamefont {A.~G.}\ \bibnamefont {White}},\ }\href@noop {} {\enquote
  {\bibinfo {title} {Communicating via ignorance},}\ } (\bibinfo {year}
  {2018}{\natexlab{b}}),\ \Eprint {http://arxiv.org/abs/1807.07383}
  {arXiv:1807.07383 [quant-ph]} \BibitemShut {NoStop}%
\bibitem [{\citenamefont {Wei}\ \emph {et~al.}(2019)\citenamefont {Wei},
  \citenamefont {Tischler}, \citenamefont {Zhao}, \citenamefont {Li},
  \citenamefont {Arrazola}, \citenamefont {Liu}, \citenamefont {Zhang},
  \citenamefont {Li}, \citenamefont {You}, \citenamefont {Wang}, \citenamefont
  {Chen}, \citenamefont {Sanders}, \citenamefont {Zhang}, \citenamefont
  {Pryde}, \citenamefont {Xu},\ and\ \citenamefont
  {Pan}}]{PhysRevLett.122.120504}%
  \BibitemOpen
  \bibfield  {author} {\bibinfo {author} {\bibfnamefont {K.}~\bibnamefont
  {Wei}}, \bibinfo {author} {\bibfnamefont {N.}~\bibnamefont {Tischler}},
  \bibinfo {author} {\bibfnamefont {S.-R.}\ \bibnamefont {Zhao}}, \bibinfo
  {author} {\bibfnamefont {Y.-H.}\ \bibnamefont {Li}}, \bibinfo {author}
  {\bibfnamefont {J.~M.}\ \bibnamefont {Arrazola}}, \bibinfo {author}
  {\bibfnamefont {Y.}~\bibnamefont {Liu}}, \bibinfo {author} {\bibfnamefont
  {W.}~\bibnamefont {Zhang}}, \bibinfo {author} {\bibfnamefont
  {H.}~\bibnamefont {Li}}, \bibinfo {author} {\bibfnamefont {L.}~\bibnamefont
  {You}}, \bibinfo {author} {\bibfnamefont {Z.}~\bibnamefont {Wang}}, \bibinfo
  {author} {\bibfnamefont {Y.-A.}\ \bibnamefont {Chen}}, \bibinfo {author}
  {\bibfnamefont {B.~C.}\ \bibnamefont {Sanders}}, \bibinfo {author}
  {\bibfnamefont {Q.}~\bibnamefont {Zhang}}, \bibinfo {author} {\bibfnamefont
  {G.~J.}\ \bibnamefont {Pryde}}, \bibinfo {author} {\bibfnamefont
  {F.}~\bibnamefont {Xu}}, \ and\ \bibinfo {author} {\bibfnamefont {J.-W.}\
  \bibnamefont {Pan}},\ }\href {\doibase 10.1103/PhysRevLett.122.120504}
  {\bibfield  {journal} {\bibinfo  {journal} {Phys. Rev. Lett.}\ }\textbf
  {\bibinfo {volume} {122}},\ \bibinfo {pages} {120504} (\bibinfo {year}
  {2019})}\BibitemShut {NoStop}%
\bibitem [{\citenamefont {Guo}\ \emph {et~al.}(2020)\citenamefont {Guo},
  \citenamefont {Hu}, \citenamefont {Hou}, \citenamefont {Cao}, \citenamefont
  {Cui}, \citenamefont {Liu}, \citenamefont {Huang}, \citenamefont {Li},
  \citenamefont {Guo},\ and\ \citenamefont
  {Chiribella}}]{PhysRevLett.124.030502}%
  \BibitemOpen
  \bibfield  {author} {\bibinfo {author} {\bibfnamefont {Y.}~\bibnamefont
  {Guo}}, \bibinfo {author} {\bibfnamefont {X.-M.}\ \bibnamefont {Hu}},
  \bibinfo {author} {\bibfnamefont {Z.-B.}\ \bibnamefont {Hou}}, \bibinfo
  {author} {\bibfnamefont {H.}~\bibnamefont {Cao}}, \bibinfo {author}
  {\bibfnamefont {J.-M.}\ \bibnamefont {Cui}}, \bibinfo {author} {\bibfnamefont
  {B.-H.}\ \bibnamefont {Liu}}, \bibinfo {author} {\bibfnamefont {Y.-F.}\
  \bibnamefont {Huang}}, \bibinfo {author} {\bibfnamefont {C.-F.}\ \bibnamefont
  {Li}}, \bibinfo {author} {\bibfnamefont {G.-C.}\ \bibnamefont {Guo}}, \ and\
  \bibinfo {author} {\bibfnamefont {G.}~\bibnamefont {Chiribella}},\ }\href
  {\doibase 10.1103/PhysRevLett.124.030502} {\bibfield  {journal} {\bibinfo
  {journal} {Phys. Rev. Lett.}\ }\textbf {\bibinfo {volume} {124}},\ \bibinfo
  {pages} {030502} (\bibinfo {year} {2020})}\BibitemShut {NoStop}%
\bibitem [{\citenamefont {Dias}(2021)}]{PhysRevA.103.012219}%
  \BibitemOpen
  \bibfield  {author} {\bibinfo {author} {\bibfnamefont {E.~O.}\ \bibnamefont
  {Dias}},\ }\href {\doibase 10.1103/PhysRevA.103.012219} {\bibfield  {journal}
  {\bibinfo  {journal} {Phys. Rev. A}\ }\textbf {\bibinfo {volume} {103}},\
  \bibinfo {pages} {012219} (\bibinfo {year} {2021})}\BibitemShut {NoStop}%
\bibitem [{\citenamefont {Peres}(1980)}]{Pe}%
  \BibitemOpen
  \bibfield  {author} {\bibinfo {author} {\bibfnamefont {A.}~\bibnamefont
  {Peres}},\ }\href {\doibase 10.1119/1.12061} {\bibfield  {journal} {\bibinfo
  {journal} {Am. J. Phys.}\ }\textbf {\bibinfo {volume} {48}},\ \bibinfo
  {pages} {552} (\bibinfo {year} {1980})}\BibitemShut {NoStop}%
\bibitem [{\citenamefont {{Jayakumar}}\ \emph {et~al.}(2014)\citenamefont
  {{Jayakumar}}, \citenamefont {{Predojevi{\'c}}}, \citenamefont {{Kauten}},
  \citenamefont {{Huber}}, \citenamefont {{Solomon}},\ and\ \citenamefont
  {{Weihs}}}]{timebinnature}%
  \BibitemOpen
  \bibfield  {author} {\bibinfo {author} {\bibfnamefont {H.}~\bibnamefont
  {{Jayakumar}}}, \bibinfo {author} {\bibfnamefont {A.}~\bibnamefont
  {{Predojevi{\'c}}}}, \bibinfo {author} {\bibfnamefont {T.}~\bibnamefont
  {{Kauten}}}, \bibinfo {author} {\bibfnamefont {T.}~\bibnamefont {{Huber}}},
  \bibinfo {author} {\bibfnamefont {G.~S.}\ \bibnamefont {{Solomon}}}, \ and\
  \bibinfo {author} {\bibfnamefont {G.}~\bibnamefont {{Weihs}}},\ }\href
  {\doibase 10.1038/ncomms5251} {\bibfield  {journal} {\bibinfo  {journal}
  {Nature Communications}\ }\textbf {\bibinfo {volume} {5}},\ \bibinfo {eid}
  {4251} (\bibinfo {year} {2014})},\ \Eprint {http://arxiv.org/abs/1305.2081}
  {arXiv:1305.2081 [quant-ph]} \BibitemShut {NoStop}%
\bibitem [{\citenamefont {Versteegh}\ \emph {et~al.}(2015)\citenamefont
  {Versteegh}, \citenamefont {Reimer}, \citenamefont {van~den Berg},
  \citenamefont {Juska}, \citenamefont {Dimastrodonato}, \citenamefont
  {Gocalinska}, \citenamefont {Pelucchi},\ and\ \citenamefont
  {Zwiller}}]{timebin}%
  \BibitemOpen
  \bibfield  {author} {\bibinfo {author} {\bibfnamefont {M.~A.~M.}\
  \bibnamefont {Versteegh}}, \bibinfo {author} {\bibfnamefont {M.~E.}\
  \bibnamefont {Reimer}}, \bibinfo {author} {\bibfnamefont {A.~A.}\
  \bibnamefont {van~den Berg}}, \bibinfo {author} {\bibfnamefont
  {G.}~\bibnamefont {Juska}}, \bibinfo {author} {\bibfnamefont
  {V.}~\bibnamefont {Dimastrodonato}}, \bibinfo {author} {\bibfnamefont
  {A.}~\bibnamefont {Gocalinska}}, \bibinfo {author} {\bibfnamefont
  {E.}~\bibnamefont {Pelucchi}}, \ and\ \bibinfo {author} {\bibfnamefont
  {V.}~\bibnamefont {Zwiller}},\ }\href {\doibase 10.1103/PhysRevA.92.033802}
  {\bibfield  {journal} {\bibinfo  {journal} {Phys. Rev. A}\ }\textbf {\bibinfo
  {volume} {92}},\ \bibinfo {pages} {033802} (\bibinfo {year}
  {2015})}\BibitemShut {NoStop}%
\bibitem [{\citenamefont {Oreshkov}(2019)}]{Oreshkov2019timedelocalized}%
  \BibitemOpen
  \bibfield  {author} {\bibinfo {author} {\bibfnamefont {O.}~\bibnamefont
  {Oreshkov}},\ }\href {\doibase 10.22331/q-2019-12-02-206} {\bibfield
  {journal} {\bibinfo  {journal} {{Quantum}}\ }\textbf {\bibinfo {volume}
  {3}},\ \bibinfo {pages} {206} (\bibinfo {year} {2019})}\BibitemShut {NoStop}%
\bibitem [{\citenamefont {Paunkovi{\'{c}}}\ and\ \citenamefont
  {Vojinovi{\'{c}}}(2020)}]{Paunkovic2020causalorders}%
  \BibitemOpen
  \bibfield  {author} {\bibinfo {author} {\bibfnamefont {N.}~\bibnamefont
  {Paunkovi{\'{c}}}}\ and\ \bibinfo {author} {\bibfnamefont {M.}~\bibnamefont
  {Vojinovi{\'{c}}}},\ }\href {\doibase 10.22331/q-2020-05-28-275} {\bibfield
  {journal} {\bibinfo  {journal} {{Quantum}}\ }\textbf {\bibinfo {volume}
  {4}},\ \bibinfo {pages} {275} (\bibinfo {year} {2020})}\BibitemShut {NoStop}%
\end{thebibliography}%
\newpage

\end{document}